\documentstyle[12pt,aaspp4]{article}

\def\etal{{et al.}}

\slugcomment{POPe-816; ApJS 128:561 (2000).}

\begin{document}


\title{The Tree--Particle--Mesh N-body Gravity Solver  }

\author{
Paul Bode\altaffilmark{\ref{Princeton}},
Jeremiah P. Ostriker\altaffilmark{\ref{Princeton}},
Guohong Xu\altaffilmark{\ref{UCSC}} 
}
\newcounter{address}
\setcounter{address}{1}
\altaffiltext{\theaddress}{Princeton University Observatory,
Princeton, NJ 08544-1001
\label{Princeton}}
\addtocounter{address}{1}
\altaffiltext{\theaddress}{University of California at Santa Cruz, Santa Cruz,
California 95064
\label{UCSC}}

\begin{abstract}
The Tree-Particle-Mesh (TPM) N-body algorithm couples the 
tree algorithm for directly computing forces on particles in an
hierarchical grouping scheme with the extremely
efficient mesh based PM structured approach. 
The combined TPM algorithm takes advantage of the fact that gravitational
forces are linear functions of the density field.
This allows the use of domain decomposition to break down the density field 
into many isolated high density regions containing a significant fraction of 
the mass but residing in a very small volume.  At low redshift, $\sim1/3$ of 
the particles in a typical large cosmological simulation can
be assigned to $\sim10^4-10^5$ separate groups occupying $\sim10^{-2.5}$ of 
the total volume.  In each of these high density regions the
gravitational potential is computed via the tree algorithm
supplemented by tidal forces from the external density distribution.
For the bulk of the volume, forces are computed via the PM
algorithm;  timesteps in this PM component are large compared to 
individually determined timesteps in the tree regions.  
Since each tree region can be treated independently,
the algorithm lends itself to very efficient parallelization using
message passing. 
We have tested the new TPM algorithm (a refinement of that
originated by Xu 1995) by comparison with results from 
Ferrell \& Bertschinger's P$^3$M code and find that, except in small
clusters, the TPM results are at least as accurate as
those obtained with the well-established P$^3$M algorithm,
while taking significantly less computing time.
Production runs of $10^9$ particles indicate that the new
code has great scientific potential when used with
distributed computing resources.
\end{abstract}

\keywords{ methods: n-body simulations --- methods: numerical 
--- cosmology: dark matter }

\section{Introduction}
In addition to the rapid increase of available computing power, 
the rise of the use of N-body simulations in astrophysics
has been driven by the development of more efficient algorithms for
evaluating the gravitational potential.
Efficient algorithms with better scaling than $\sim N^2 $ take two
general forms.  First, one can introduce a rectilinear spatial grid and, 
taking advantage of Fast Fourier Transforms, solve
Poisson's equation on this grid in Fourier space--- the well-known
Particle--Mesh (PM) method,  which, while very fast, limits the spatial
resolution to the grid spacing.  
To gain finer resolution one can introduce smaller subgrids
(e.g. the ART code of \cite{Kravetal97}; see also \cite{NormBrya99});
alternatively one can compute the short-range
interactions directly (the Particle--Particle--Particle--Mesh,
or P$^3$M method (\cite{Efst85}; \cite{FerrBert94})).  
One widely used code (AP$^3$M) combines both of these refinements
(\cite{Couc91}; \cite{PearCouc97}). 
The second general approach is to approximate
long-range interactions which are less important to an accurate
determination of the force, by grouping together distant particles.
These are known as Tree methods since a tree data structure is
used to hold the moments of the mass distribution in nested subvolumes
(\cite{BarnHut86}; \cite{Hern87}). 
ART and AP$^3$M are discussed in more detail by by \cite{Knebetal00};
for a review of the field see \cite{Couc97}.

All of these algorithms are more difficult to implement on
parallel computers with distributed memory than on single processor
machines.  Gravity acts over
long scales and gravitational collapse creates highly inhomogeneous
spatial distributions, yet with parallel computers one needs to limit 
the amount of communication and give different processors roughly
equal computing loads.   The problem is one of domain decomposition---
locating spatially compact regions and deciding which data is needed
to find the potential within that region.

\cite{Xu95} introduced a new N-body gravity solver which deals
with this problem in a natural way.
The Tree--Particle--Mesh (TPM) 
approach is similar to the P$^3$M method, in that the
the long-range force is handled by a PM code and the
short-range force is handled by a different method---  in this case
using a tree code,  with the key difference that the tree code is used
in adaptively determined regions of arbitrary geometry.
In this paper we describe several improvements
to the TPM code, and compare the results with those obtained by
the P$^3$M method.
Our goal was to improve and to test the new algorithm while 
designing an implementation that could be parallelized 
efficiently and was optimal for use as a coarse grained
method suitable for distributed computational architectures,
including those having large latency.
Section 2 describes the method, Section 3 the
basis (density threshold) for domain decomposition, Section
4 the parallelism of the implemented algorithm (using
message passing), and Section 5 tests and compares with the
well calibrated P$^3$M algorithm.

The implementation presented in this paper is oriented towards a specific 
cosmological problem-- the formation of large clusters-- and we will be 
discussing it in that setting.  However, this algorithm could be used for 
many particle simulation applications, both in astrophysics and other 
fields;  it should be beneficial in situations where the density distribution 
allows one to divide the particles into many isolated groups.  
Thus we will conclude this section with a brief summary of the specific
cosmological context for those unfamiliar with it.
A large cubical volume is simulated with periodic boundary conditions.
The simulation begins in the linear regime;
particles are displaced slightly from a uniform grid,
giving Gaussian perturbations to a nearly constant density field.
The particles are followed as they move under their mutual gravitational
attraction.  Over time, gravitational instability
causes the initially small overdensities to collapse, forming highly
dense halos (with central densities a factor of $\sim10^5$ higher than
the average).  These halos are distributed along filaments 
surrounding large, low-density voids.
The TPM algorithm was developed to deal with
this highly inhomogeneous structure.

\section{The TPM algorithm}

The basic idea behind the TPM algorithm is to identify dense regions
and use a tree code to evolve them; low density regions and all 
long--range interactions are handled by a PM code.  A general
outline of the algorithm is:

\begin{enumerate}
\item  Find the total density on a grid.
\item  Based on the grid density, decompose the volume into a 
background PM volume and a large number of isolated high density 
regions.  Every particle is then assigned to either the PM background
or a specific tree.
\item  Integrate the motion of the PM particles (those not in any tree)
        using the PM gravitational potential computed on the grid.
\item  For each tree in turn integrate the motion of the particles,
        using a smaller time step if needed;
	forces internal to the tree are found with a tree algorithm
	(\cite{Hern87}), added to the tidal forces from the
	external mass distribution taken from the PM grid.
\item  step global time forward, go back to step 1.
\end{enumerate}

In this section we will consider certain aspects of this process
in detail, and conclude with a more complete outline of the algorithm.

\subsection{Spatial Decomposition} \label{sectreeass}

We wish to locate regions of interest which will be treated with
greater resolution in both space and time;  for the purposes of
cosmological structure formation this translates into regions of 
high density.  It also is necessary that these regions remain
physically distinct during the long PM time step (determined by
the Courant condition) so that the 
mesh-based code accurately handles interactions between two such regions. 
The process we use can be thought of as finding regions enclosed by an
isodensity contour.  If one imagines the isodensity contours through a typical
simulation at some density threshold
$\rho_{\rm thr}>\bar{\rho}$, space is divided
into a large number of typically isolated regions with 
$\rho>\rho_{\rm thr}$ plus a multiply connected low density background
filling most of the volume.

To locate isolated, dense regions we begin with the grid density,
which has been calculated already by the PM part of the code.
Each grid cell which is above a given threshold density $\rho_{\rm thr}$ is
identified and given a unique positive integer key (the choice of
$\rho_{\rm thr}$ is discussed in Section~\ref{secdenth}).
Cells are then grouped by a `friends-of-friends' approach: for
each cell with a nonzero key the 26 neighboring cells are examined and,
if two adjacent cells are both above the threshold, they are
grouped together by making their keys identical.  The end result
is isolated groups of cells, each separated from the other groups
by at least one cell.  If a wider separation between these regions 
is desired, one can examine a larger number of neighboring cells.
The method is ``unstructured'' in the sense that the geometry of
each region is not specified in advance, except insofar as it is
singly connected.  The shape of the region can be spheroidal, planar,
or filamentary as needed.
 
To assign particles to trees, the process used to find
the density on the grid (described in the next section)
is repeated.  This involves locating the grid
cell to which some portion of a particle's mass is to be added, 
so it is easy to check
at the same time whether that cell has a nonzero key and,
if it does, to add that particle to the appropriate tree.  Thus any
particle that contributes mass to a cell with density above
the threshold is put in a tree.  Because of the spatial separation
of the active regions (they are buffered by at least one non-tree
cell) a particle will only belong to one tree even
though it contributes mass to more than one cell.
 
An example of this in practice is shown in Figure~\ref{figfind}.
In the bottom panel, all particles in a small piece of a larger
simulation are shown in projection.
The grid and the location of active cells
are shown in the top panel; each isolated region is indicated
by a unique numerical key.  In a couple of cases it appears that
different regions are in adjacent cells, but in fact they are
separated in the third dimension-- the region shown is 10 cells thick.
In the lower of the middle two panels, the particles assigned to trees 
are shown with different symbols indicating membership in different trees.
In the other panel the residual PM particle positions are plotted, 
demonstrating their much lower density contrast as compared to those
in trees.

\subsection{Force Decomposition} \label{secforce}

As in \cite{Xu95}, the force is decomposed into that which is internal
to the tree and that due to all other mass:
\begin{equation}
{\bf F} = {\bf F}_{\rm internal} + {\bf F}_{\rm external}.
\end{equation}
However, we do this in a different manner, described in this
section, than was done in \cite{Xu95}.

The first step in obtaining the particle accelerations is to 
obtain the PM gravitational potential.  The masses $m_p$ of the $N$ 
particles (including those in trees) are assigned to the grid cells using
CIC weighting:
\begin{mathletters} \label{eqncic}
\begin{eqnarray}
\rho_{\rm all}(i,j,k) = \sum\limits_{p=1}^{N} m_p w_i w_j w_k , \\
w_i = \cases{
	 1 - |x_p - i| & for $|x_p -i|<1 ,$ \cr
	 0 & otherwise, }
\end{eqnarray}
\end{mathletters}
where $x_p$ is a particle's $x$ coordinate in units where the
grid spacing is unity.
The potential $\Phi_{\rm PM,all}$, assuming periodic boundary conditions, 
is then found by solving Poisson's equation using the standard FFT 
technique (Hockney \& Eastwood 1981).

Once a tree has been identified, we wish to know the forces from all the
mass not included in that tree; thus the contribution of the tree
itself must be removed from the global potential. This step will have
to be done for each tree in turn.
The density is found exactly as before, except this time summing over only 
the particles in the tree:
\begin{equation}
\rho_{\rm tree}(i,j,k) = \sum\limits_{tree} m_p w_i w_j w_k  \\
\end{equation}
Using this density, we solve Poisson's equation again, except that
non-periodic boundary conditions are used (Hockney \& Eastwood 1981).
The resulting potential $\Phi_{\rm NP,tree}$ is the contribution
that the tree made to $\Phi_{\rm PM,all}$ without counting the
ghost images due to the periodic boundary conditions of the latter.
The force on a tree particle exerted by all the mass not in the
tree (including the periodic copies of the tree) is then
\begin{equation}
{\bf F}_{\rm external} = 
       \sum_{i,j,k} w_i w_j w_k \nabla \Phi _{\rm PM,all} -
       \sum_{i,j,k} w_i w_j w_k \nabla \Phi _{\rm NP,tree} 
\end{equation}
Thus tidal forces within a tree region are computed on the
mesh scale in a consistent manner, with interpolation used
as required to find the forces on individual particles.

Calculating the non-periodic potential with FFTs involves using
a grid which is eight times larger in volume than 
that containing the actual mesh of
interest,  but since trees are compact and isolated regions, the
volume of the larger grid which is non-zero is quite small.
Thus the FFT which is computed for each tree can be done on a
smaller grid as long as the grid spacing remains the same as for
the larger periodic FFT;  we do this by embedding the irregular
tree region in a cubic subgrid, padding with empty cells 
as needed.

The final step is to calculate the internal forces
${\bf F}_{\rm internal}$  for each tree.
We do this with the tree code of Hernquist (1987).  
Since the periodic nature of the potential was taken into
account in finding the external forces, no Ewald summation
is needed.
Time stepping is handled in the same manner as \cite{Xu95}.
That is, the PM potential is determined at the center of
the large PM timestep, and each tree has its own, possibly
smaller, timestep.
There are a couple of slight differences:  in Equation 15 of
\cite{Xu95} we use the parameter $\beta=0.05$, and we decrease
$\delta t_{TREE}$ so that 97.5\% of the tree particles
satisfy $\delta t_i\geq \delta t_{TREE}$.

\subsection{Detailed Outline}

To sum up this section we give a more detailed outline of the code.
All particles begin with the same time step 
$\Delta t=\Delta t_{PM}$;
the velocities are given at time $t$ and the positions at time
$t+\Delta t/2$ (as described in Xu 1995).

\begin{enumerate}
\item Using the density from the previous step, we identify all
particles belonging to trees, and to which tree (if any)
each particle belongs (Section~\ref{sectreeass}).

\item The time step for each tree is computed,
and particle positions are adjusted if $\Delta t$ has
changed for that particle (Hernquist \& Katz 1989). 
This can occur if a particle
joins or leaves a tree, of if the tree time step has changed.

\item The total density due to all particles at time $t+\Delta t_{PM}/2$
is found on a grid using Equation~\ref{eqncic}.
The potential $\Phi_{\rm PM,all}$ is found
from this density, and the PM acceleration at mid-step is found
for each particle. 

\item Each tree is then dealt with in turn. First, the tree contribution
to the PM acceleration is removed, as described in 
Section~\ref{secforce}.
Next the tree is stepped forward with a smaller time step
using the tree code of
Hernquist (1987), with the external forces included.

\item All particles not in trees are stepped forward using
the PM acceleration.  The global time and cosmological parameters
are updated, completing the step.
\end{enumerate}

\section{The Density Threshold} \label{secdenth}
In Section~\ref{sectreeass} the threshold density $\rho_{\rm thr}$ 
was introduced to demarcate dense regions which would be followed
with higher resolution.   
The best choice of this parameter depends
on a number of considerations.  One could set $\rho_{\rm thr}$ to
be such a low value that nearly all particles are in trees, or that
only one large tree exists, thereby destroying the efficiency that
the TPM algorithm is designed to give.  On the other hand, too high
a value would leave many interesting regions computed
at the low resolution of the PM code.
When modeling gravitational instability, one must also keep in mind
that the density evolves from having only small overdensities initially
to a state where there are a few regions of very large overdensity;
thus the ideal threshold will evolve with time.
With these considerations in mind,
we base $\rho_{\rm thr}$ on the grid density as: 
\begin{equation} \label{eqnrhothr}
\rho_{\rm thr} = A\bar{\rho}+B\sigma
\end{equation}
where $\bar{\rho}$ is the mean density in a cell, and $\sigma$ is the
standard deviation of the cell densities.  
With this equation, the first two moments of the density distribution
are used to fix $\rho_{\rm thr}$ in an adaptive manner.
The coefficient $A$ is set to prevent the selection of too many or too large
trees when $\sigma$ is small; its value will be near unity.  The choice of
$B$ will determine what fraction of particles will be placed in trees
when $\sigma$ is large.  This choice depends on the parameters of the
simulation such as the cosmological model (including the choice of $\sigma_8$)
and the size of a grid cell.
We choose a value of $B$ which will place $\sim 1/3$ of the particles 
in trees at the end of the simulation.

Figure~\ref{fighaloh} shows how tree properties vary over the course
of a large LCDM simulation, using $A=0.9$ and $B=4.0$ in
Equation~\ref{eqnrhothr}.
The value of $\sigma$ begins at 0.1, so at high redshift 
$\rho_{\rm thr}\lesssim1.5\bar{\rho}$.
This leads to a large number of trees which are low in mass
and diffuse.  As time goes on, these slight overdensities
collapse and merge together, resulting in denser concentrations
of mass. Also, $\sigma$ becomes larger (increasing to 4.1 by the
end of the simulation), so a larger concentration of mass is needed
before a region is identified as a tree. Thus the original
distribution of trees evolves into one with fewer trees, but
at higher masses (though at any given time the masses of trees roughly
follow a power law distribution).
The typical volume within tree regions also increases with time, but
the total volume covered by trees (measured by the number of cells
above $\rho_{\rm thr}$) decreases.  Given the roughly log--normal
distribution of density resulting from gravitational instability,
the total volume in tree regions is less than one percent even
when they contain $\sim 30$\% of the mass.
The rise in $\rho_{\rm thr}$ means that the size of the smallest tree 
found also rises-- from 4 to 40 particles over the course of this run. 
This raises an issue that must be noted when understanding
the results of a TPM run: the choice of $\rho_{\rm thr}$ introduces
a minimum size below which the results are no better than in a PM code.
This is discussed in more detail in Section 5.

\section{Parallelism}
One of the strengths of the TPM algorithm is that after the PM step,
each tree presents a self-contained problem:  given the particle positions,
velocities, and tidal forces, the tree stepping can be completed without
the need to access any other data, since  
the effect of the outside
universe is summarized by the tidal forces in the small tree region.
This makes the tree part of the code intrinsically parallel.
What makes such a separation possible is that during the multiple
timesteps required to integrate particle orbits within a 
dense tree region the tidal forces may be deemed constant; the code
is self-consistent in that the density on the PM grid is only
determined on the Courant timescale for that particle distribution.

Our parallel implementation of the TPM method uses a distributed
memory model and the MPI message
passing library, in order to maximize the portability of the
code.  The PM portion of the code is made parallel in a manner
similar to that described in \cite{BodeXuCe96}.  This scales well,
and takes a small fraction of the total time as compared to the
tree portion of the code.

Two steps are made to ensure load balancing the tree part of the code.
First, trees are distributed among processors in a manner intended to
equalize the amount of work done.  The time it takes for a particular
tree to be computed depends on the size of the tree, the cost of
computing the force scaling roughly
as $N\log{N}$.  
As trees are
assigned to processors, a running tally is kept of the amount of work
given to each node,  and the largest unassigned tree is assigned to the
processor given the least amount of work.
The tree particles are then distributed among the processors, and each
processor deals with its assigned trees, moving from the largest to the
smallest.   There is also a dynamic component to the load balancing:
when a node has completed all of its assigned trees, it queries another
process to see if that one is also finished.  If that process still has 
an uncomputed tree remaining 
in its own list, it sends all the necessary tree data to the querying node. 
That node then evolves the tree and sends the final state back to the
node that had the tree originally.  Thus nodes that finish earlier than
expected do not remain idle.

The scaling of the code is shown for two different size problems in
Figure~\ref{figtiming}; the times shown are for when
the underlying LCDM model is at low redshift (z=0.5),
meaning that clustering is significant and calculating tree forces
dominates the CPU time.  At higher redshift, when the trees are 
less massive and more diffuse, the timing would be more like
that of a PM code (this can be seen from Table~\ref{tbl-1}).
These timing tests were run on an SGI Origin 2000 with 250 MHz chips;
the scaling on a PC cluster with a fast interconnect was found to be
quite similar.
The $512^3$ model is the one shown in Figure~\ref{fighaloh}; 
it scales reasonably well up to the largest $NPE$ we
attempted; compared to $NPE=32$, the efficiency is better than 90\%
at $NPE=128$, and 80\% at $NPE=256$.
When using 32 nodes the code required 512 Mbyte per node,
so we did not try any smaller runs.
The $256^3$ times are for the same LCDM model except with a
smaller box size (150 Mpc/$h$) and $\rho_{\rm thr} = 0.85\bar{\rho}+4.0\sigma$.
Since the largest nonlinear scale is a larger fraction of the
box size,  a greater fraction of particles (37\%) are placed in trees
and the largest tree contains a greater proportion of the mass.
This $256^3$ model scales extremely well from 4 to 16 processors,
but drops to 70\% efficiency at 32 nodes, and beyond 64 nodes does
not speed up at all.  The reason for this is that the largest
tree in this simulation contains one percent of all particles, 
which means this one tree takes a few percent of the entire CPU
time devoted to trees.  As $NPE$ is increased, the time it takes
to complete this one tree becomes the major part of the total time.
The solution to this problem is to allow more than one processor
to work on the same tree, which is quite possible (e.g. Dav\'e et al.
1997 and the references therein; see also Xu 1995).  

The division of the total time between different components of
the code is shown in Table~\ref{tbl-1} for both low and high
redshift.  At low redshift the tree calculations dominate the
total time (as long as this part of the code is load balanced--
the rise in overhead for the $256^3$ model when $NPE\geq 32$ is
due to imbalance, as discussed above).
At high redshift the trees are smaller, so the overhead related
to domain decomposition takes a large fraction of the total time;
the main difference between the two redshifts is the rising cost
of the tree calculations as trees become more massive and require
more timesteps. 
Comparison with the the P$^3$M code of Ferrell \& Bertschinger (1994)
(made parallel by \cite{Fred97}) shows that TPM (with 30\% of the
particles in trees) takes slightly less
time than P$^3$M if all the trees keep to the PM time step.  
Allowing trees to have individual time steps speeds up the TPM code
by a factor of three to four.
In the present implementation, particles within
the same tree all use the same timestep;  implementing multiple
time steps within trees could further save a significant
amount of computer time (roughly another factor of three)
without loss of accuracy.

The memory per process used by our current implementation is $20N/NPE$
reals when there is one cell per particle. 
This includes for each particle $\vec{x}, \vec{v}, \vec{a}$, and
three integer quantities (a particle ID number, a tree membership
key, and the number of steps per PM step).  The remaining space
is used by the mesh part of the code, and reused as temporary
storage during the tree stepping.
Because the grid density from the previous step is saved, the memory
used could be reduced to $17N/NPE$ at the cost of computing the
density twice per step.

The $1024^3$ point shown in Figure~\ref{figtiming} is for the same
cosmological model and box size as the $512^3$ run, but with eight 
times as many particles.
This run shows the great potential of the TPM algorithm.  At lower
redshifts over 80\% of the computational time is spent finding tree
forces--- precisely that portion of the code which involves no
communication;  thus a run of this size would be able to efficiently
utilize even more processors.  This does not necessarily mean using
a larger supercomputer; rather, one could use networked PC's or
workstations.  These distributed resources could be used to receive
a single tree or small group of trees, do the required time stepping
in isolation, and send back the final state.  The time required to
evolve a single tree varies from less than a second to a couple
minutes, so even in situations with a high network latency the cost
of message passing need not be prohibitive.

\section{Tests of the Code}

To test how the code performs in a standard cosmological simulation
we ran both TPM and the P$^3$M code
of Ferrell \& Bertschinger (1994) with the
same initial conditions.  The test case contains $128^3$ particles in a
150 Mpc/$h$ box, with a flat LCDM cosmological model close to that
proposed by \cite{OstrStei95}: 
$\Omega_m=0.37$, $\Lambda=0.63$, $H_o=70$ km/s/Mpc, $\sigma_8=0.8$,
and tilt $n=0.95$.  The softening length of the particles is 
$\epsilon=18.31$ kpc/$h$.  The number of mesh points in the PM grid
was $256^3$ for the P$^3$M run and $128^3$ for TPM.  The 
TPM threshold density was 
$\rho_{\rm thr}=0.85\bar{\rho}+4.0\sigma^2$, so a third of the
particles were contained in trees by $z=0$.
In the tree code an opening angle of $\theta=0.5$ was used.

Figure~\ref{figsnaps} shows projected particle positions at the 
final redshift $z=0$ for a portion of the volume around the largest 
halo that had formed.  One important difference between the two
codes can be seen by examining this figure.  It is clear that 
the largest structures are quite similar in both cases; but
notice that a number of small halos can be identified in the
P$^3$M snapshot that are not present in TPM.  
To verify this visual appearance in a more quantitative manner,
bound halos were identified with DENMAX (\cite{GelbBert94}).
The resulting mass functions for the two codes are shown in
Figure~\ref{figcnofm}.  The agreement is good for trees with
more than 100 particles, but the TPM model has fewer small
halos with less than 100 particles, confirming the visual impression.

The cause of this difference arises from the choice of $\rho_{\rm thr}$.
Those objects that collapse early, which through merger and accretion
will end up having higher masses, are identified when only slightly
overdense and thus are followed at higher resolution throughout their
formation.  As $\rho_{\rm thr}$ rises, a halo must reach a higher
overdensity before being followed with the tree code, so objects
which collapse at late times are simulated at lower resolution. In
this test case, the smallest tree at $z=0$ contains 66 particles,
so it is unsurprising that TPM has fewer halos near and below
this size.

This effect is shown in a different way in Figure~\ref{figcorrel},
where the two-point correlation function $\xi(r_{12})$ is shown for 
the two test runs.  For separations $r_{12}>1$Mpc 
there is no discernible difference between the P$^3$M  and
TPM particle correlations.
However, when all particles are included in calculating $\xi$, 
the P$^3$M code yields a greater correlation at smaller
separations.  We also selected from each simulation the 
particles contained in the 1000 largest halos found by DENMAX, 
and redid the calculation with only those particles. In this case,
the TPM correlation function is the same as the P$^3$M, and in
fact is higher for $r_{12}<10\epsilon$.
This demonstrates clearly that the lower TPM correlation function
in the former case is an effect of the higher force resolution
of P$^3$M in small halos and other regions where $\rho<\rho_{\rm thr}$.
Within TPM halos followed as trees, the resolution is as good 
as (or better than) in P$^3$M; the
difference in $\xi$ computed for halo particles only is 
most likely due to differences in 
softening (the tree code uses a spline kernel while P$^3$M uses
a Plummer law) and in the time stepping.

The distribution of velocities is also sensitive to resolution effects.
To examine this, particle pairs were divided into 30 logarithmically spaced 
bins, with bin centers between 50 kpc and 20 Mpc;
for each pair the line-of-sight velocity difference $v_{12}$ was computed.
Histograms showing the distribution of $v_{12}$ in selected radial bins
are shown in Figure~\ref{figvhist}.  If only particles in the 100
largest halos are considered, the two codes are indistinguishable.
But again, a difference becomes noticeable as more particles are 
included --- the P$^3$M halos begin to show more pairs with a 
small velocity difference ($v_{12}<250$km/s).  Since the P$^3$M code
is following smaller halos with higher resolution, these halos
have smaller cores and a cooler velocity distribution than TPM halos
with the same mass.

In order to compare the properties of individual collapsed objects,
we selected a group of halos as follows.
First, we chose those DENMAX halos without a more massive neighbor
within 2 Mpc/$h$. 
The spherically averaged density profile $\rho(r)$ was found for
each halo, and a fit to the NFW profile 
(Navarro, Frenk \& White 1997) was computed by
a $\chi^2$ minimization; those with less than 99.5\% likelihood
were excluded from further analysis.
This fitting procedure  repositioned the centers onto the densest
region of the halo;  we removed those halos where  the positions 
found in the two models differed by more than $r_{200}/3$, 
in order to be sure that it is the same halo being examined in both cases.
Figure~\ref{fighaloprof} shows 
the $\rho(r)$ for a few halos 
selected in this manner; the agreement is quite good, and
within statistically expected fluctuations.
If the TPM code had a lower resolution then a broader halo profile with a
lower density peak would result, but this is not seen.

Comparisons of other derived properties are shown in 
Figure~\ref{fighalocomp}.  In each case we plot the fractional
difference of the two models: [f(TPM)-f(P$^3$M)]/0.5[f(TPM)+f(P$^3$M)].
The top panel shows the number of particles within 1.5 Mpc/$h$ of 
the center, and the second panel shows 
the velocity dispersion.
The agreement in both cases is good-- the dispersion is 7\% and
9\% respectively, with no systematic offset or discernible
trend with halo mass.
The third panel compares $r_{200}$ from the NFW fits, which also
agrees quite well, the dispersion being 5\%. At the low mass end 
there are some TPM halos
with sizes more than 20\% larger, but these
are also the ones with the smallest  $r_{200}$.
The final panel compares the core radius $r_s$ resulting from the
NFW profile fits, which shows the most variation between codes.
There are a number of TPM halos with substantially larger cores
(particularly at low mass), but the average TPM core size is
smaller by 10\% than that in P$^3$M.  
It appears that most TPM cores have in general
been followed with the same or higher resolution than obtained with the
P$^3$M code, but a few have not.
Examination of those halos with largest differences often show
substructure or high ellipticity, but this is not always the case.

\section{Summary}
In the current environment, those wishing to carry out high
resolution simulations must tailor their approach to exploit
parallel and distributed computing architectures. In this paper
we have presented an algorithm for evolving cosmological
structure formation which is well suited to such machines.
By suitable domain decomposition, one large volume is broken
up into a large number of smaller regions, each of which
can be solved in isolation.  This simplifies balancing the load between 
different processes, and makes it possible to use machines with high 
latency (e.g. a large number of physically
distributed workstations) efficiently.  Furthermore, it ensures
that higher resolution in both space and time is applied in only 
those regions which require it.

An important parameter in the TPM code is the density threshold.
By tying this parameter to the first and second moments of the
density distribution, it is possible
to follow initially small overdensities as they collapse and
thus simulate halo evolution with as high resolution as the
more common P$^3$M code.  However, it is best to consider only
those halos which contain twice as many particles as the smallest
tree.
Recently \cite{Bagla99} introduced a different method of combining
gridded and tree codes called TreePM. This algorithm computes both 
a PM and a tree force for every particle,  which has the advantage
of uniform resolution for all particles.  The performance of TPM
in lower density regions can always be improved by lowering the
density threshold, though this may lead to unacceptably large
trees.  Another possibility which we intend to investigate,
is to create a ``TP$^3$M'', which
uses P$^3$M rather than PM in the non-tree volume.  This could
be quite practicable, since the particle-particle interactions are
not expensive to compute when the density is low.

However, it may be that increased force resolution in low density
regions is not a true improvement.  \cite{Meloetal1997} and
\cite{Splinetal1998} showed that discreteness and two-body scattering
effects become problematic when the force resolution outstrips
the corresponding mass resolution.  This led to a recent investigation
by \cite{Knebetal00}, who concluded that strong two-body scattering
can lead to numerical effects, particularly when the local 
interparticle separation is large or the time step is too long;
slowly moving pairs of particles may suffer interactions which do
not conserve energy.  The TPM code will be less prone to such
effects because low density regions use lower force resolution;
only as the local mass resolution increases does the force resolution
become higher, and simultaneously the time step will tend to become smaller.

This research was supported by NSF Grants AST-9318185 and
AST-9803137 (under Subgrant 99-184), and the NCSA Grand Challenge 
Computation Cosmology Partnership under NSF
Cooperative Agreement ACI-9619019, PACI Subaward 766.
Many thanks to to Ed Bertschinger for use of his P$^3$M code,
and Lars Hernquist for supplying a copy of his tree code.

\clearpage

\epsscale{0.80}
\plotone{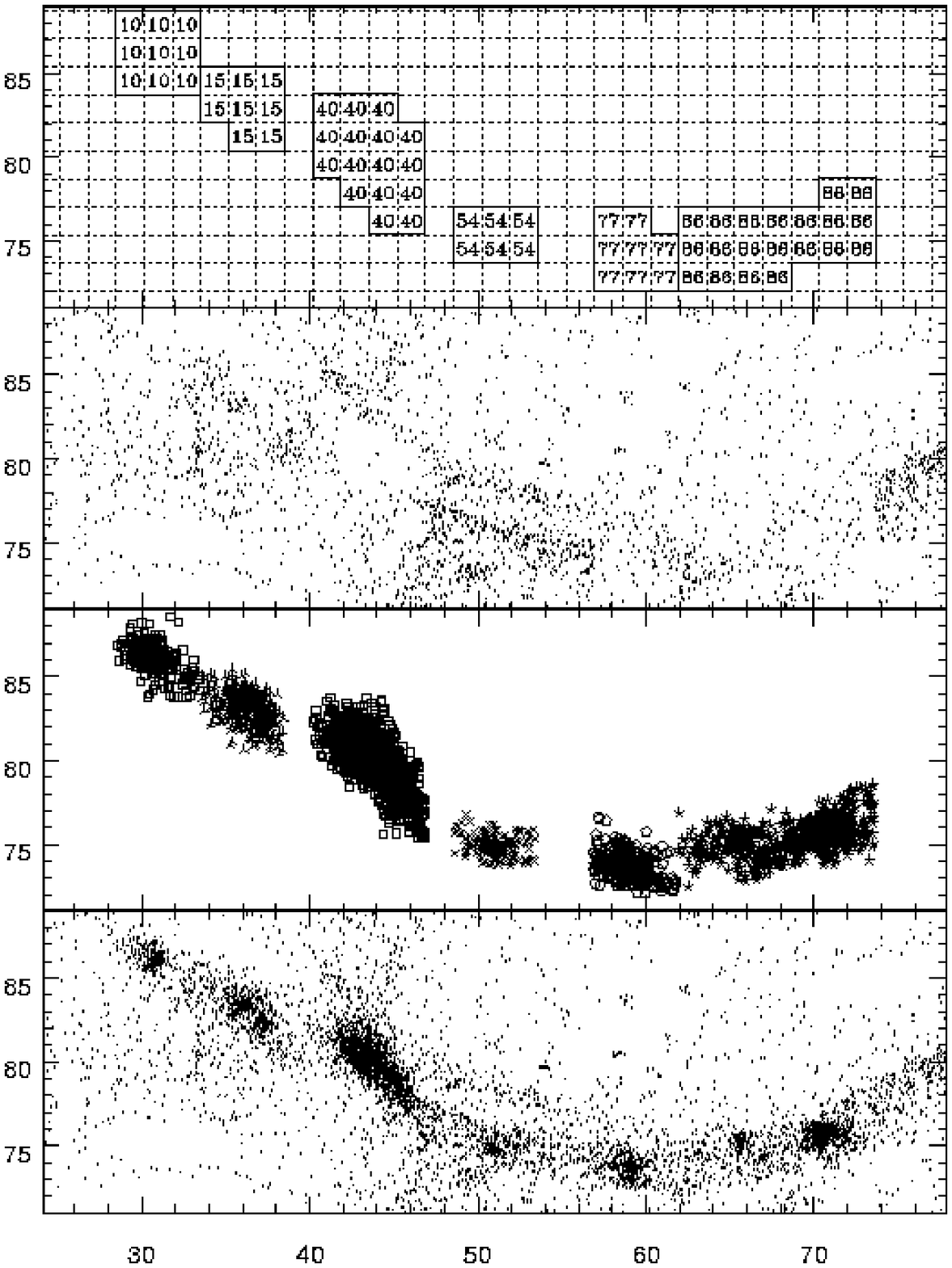}
\figcaption[f1.eps]{Defining 
trees.  From bottom to top:  all particles;  
tree particles, with different symbols indicate
membership in different trees; PM particles; and tree regions.
In the top panel, dotted lines show the mesh spacing and numbers indicate
active cells.  
Axis labels are in Mpc; this volume is drawn from a larger
simulation and is 10 Mpc thick. The threshold density is 
$20\bar{\rho}$.   The apparent adjacency of some regions shown in 
the top panel is
a projection effect-- all tree regions are in fact spatially distinct.
\label{figfind} }
\clearpage

\epsscale{1.0}
\plotone{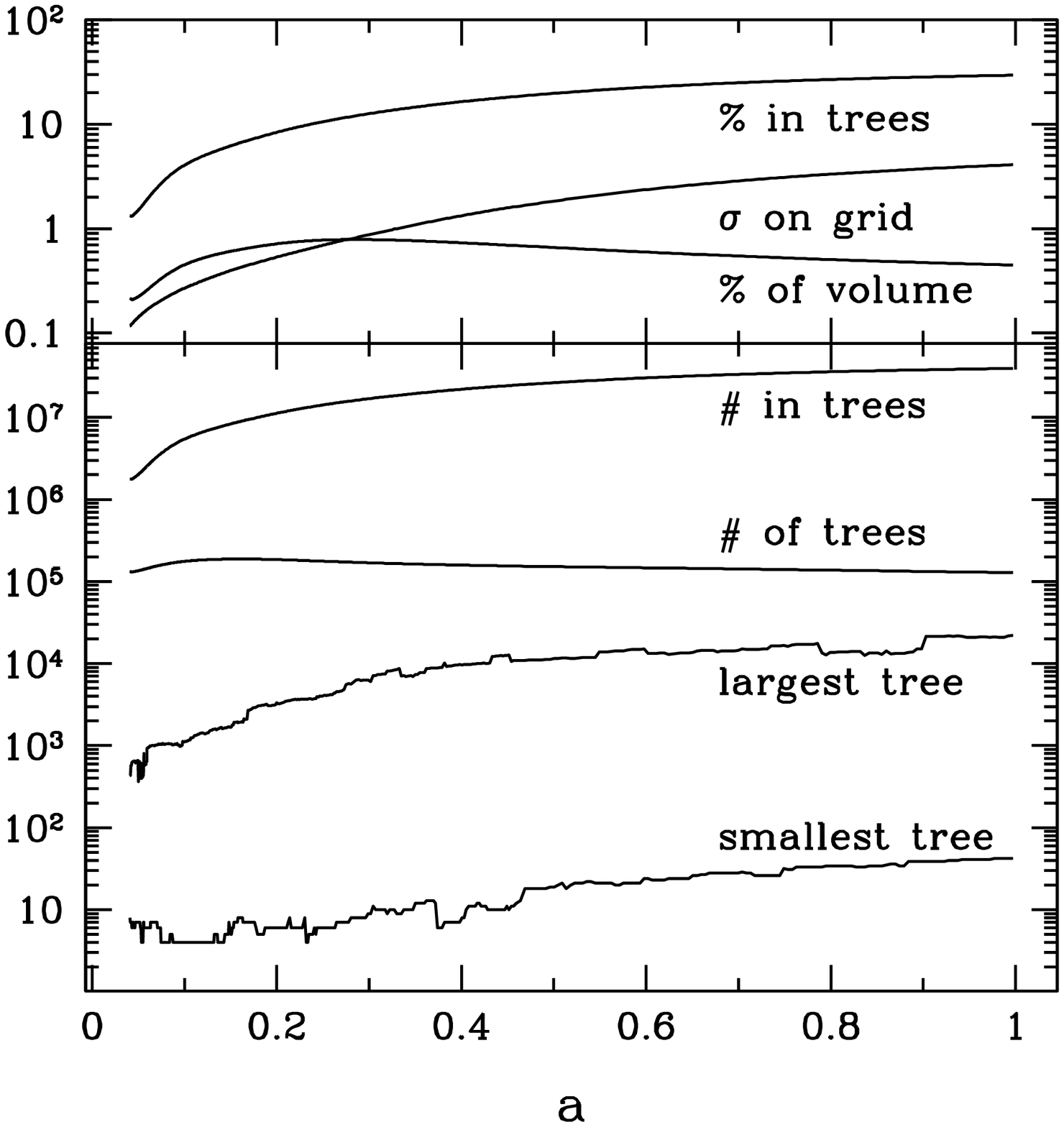}
\figcaption[f2.eps]{Properties of trees as a function of
expansion factor $a$, from $a=0.04$ to 1. 
From top to bottom: the percentage of all particles in trees; 
the standard deviation of the density of the grid cells
(in units where the mean $\bar{\rho}=1$);
the percentage of the total volume occupied by trees (measured
as the number of active cells divided by the total number of cells);
the number of particles in trees; the number of separate
trees; the number of particles in the largest tree; and the number in
the smallest tree. 
The model is an $N=512^3=10^{8.1}$ 
LCDM simulation of a 1000 Mpc/$h$ box; the 
threshold density is $\rho_{\rm thr} = 0.9\bar{\rho}+4.0\sigma$.
\label{fighaloh} }

\plotone{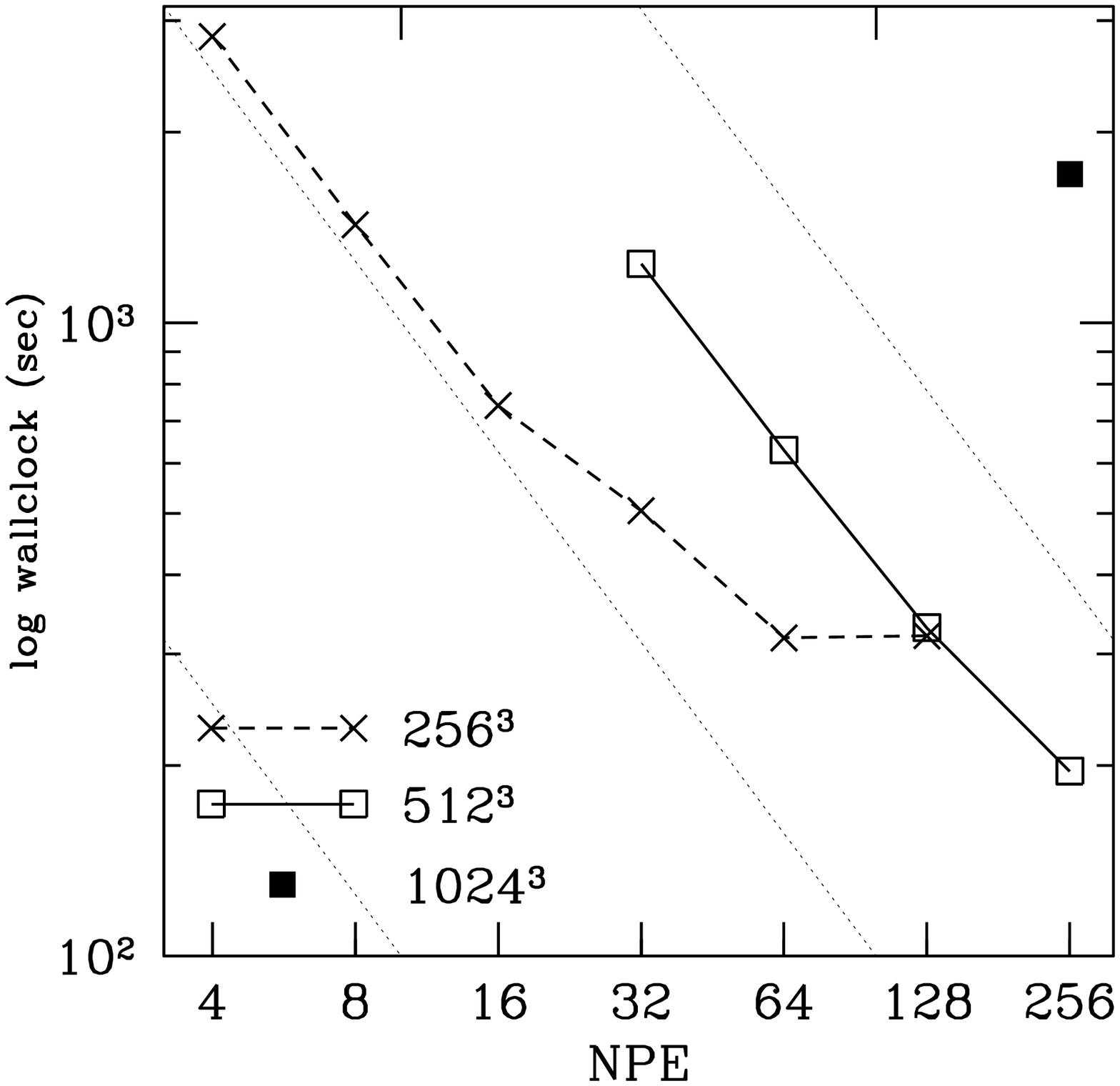}
\figcaption[f3.eps]{Timing as a function of the number of
processors, for models at $z=0.5$.  The labels give the number of
particles, which equals the number of cells in each case.
The thin dotted lines show the slope expected for a perfect 
scaling of $\sim NCPU^{-1}$.  The time shown is for one PM step;
the number of steps for individual trees varies, up to 10 steps
for the larger ones.
\label{figtiming} }

\epsscale{0.9}
\plotone{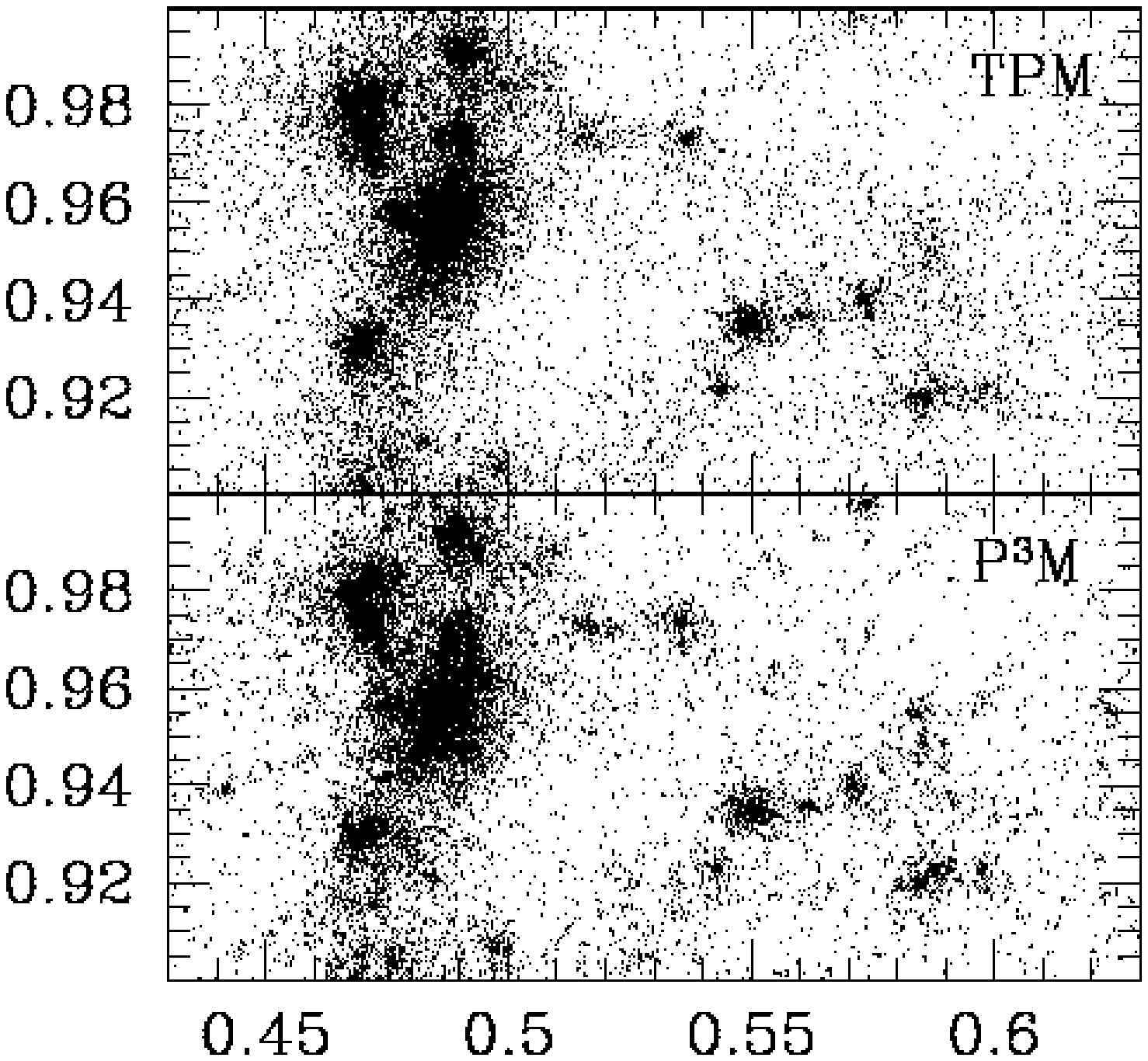}
\figcaption[f4.eps]{Final particle positions near a large
halo formed in a $N=128^3$ LCDM simulation, run with TPM {\it (top)}
and P$^3$M {\it (bottom)}.
\label{figsnaps} }

\epsscale{0.9}
\plotone{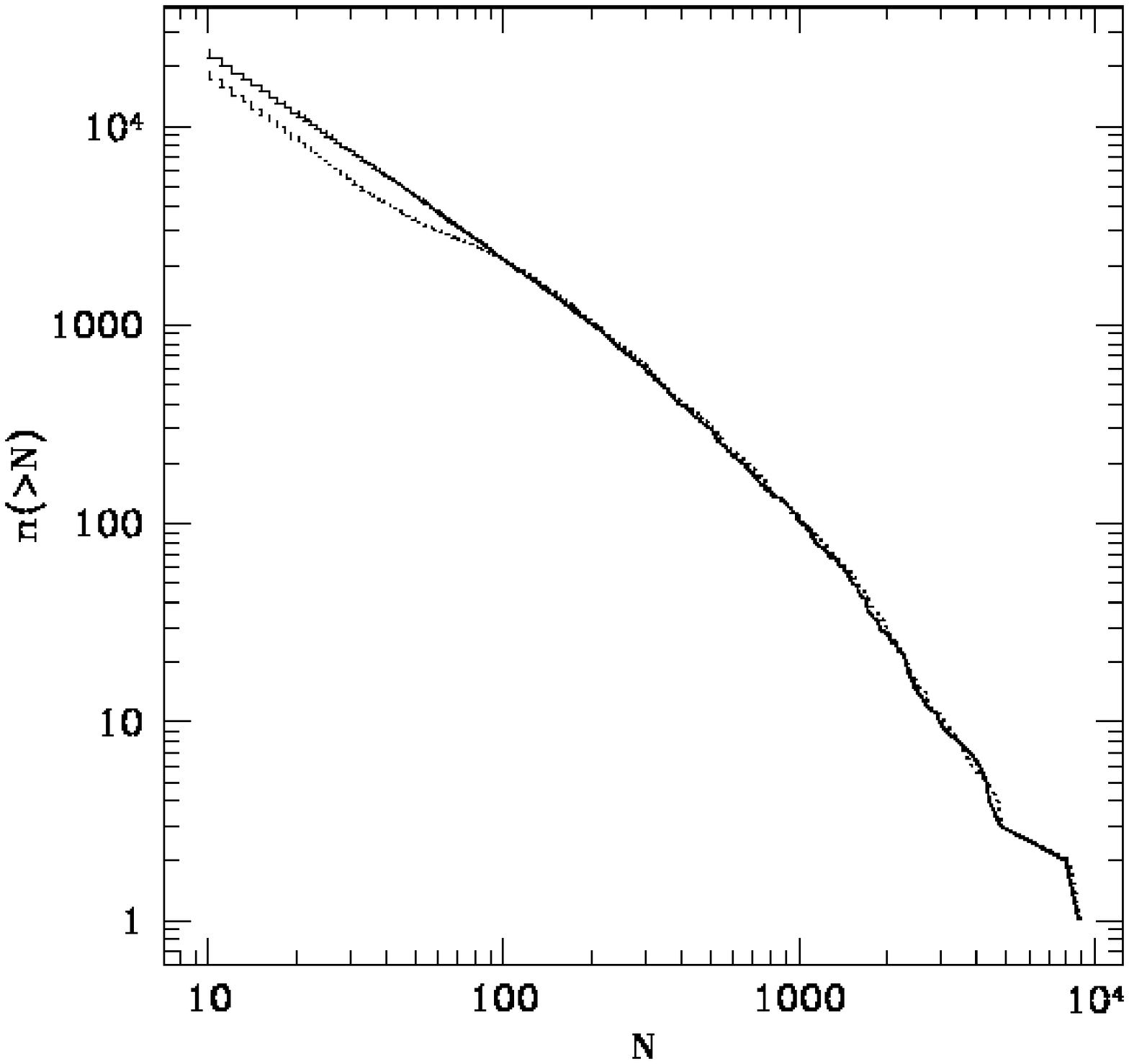}
\figcaption[f5.eps]{The mass function (shown as number of
halos containing more than $N$ particles) resulting from the
TPM code {\it (dashed)} and the P$^3$M code {\it (solid)}. Details
of the simulation are described in the text.
\label{figcnofm} }

\epsscale{1.0}
\plotone{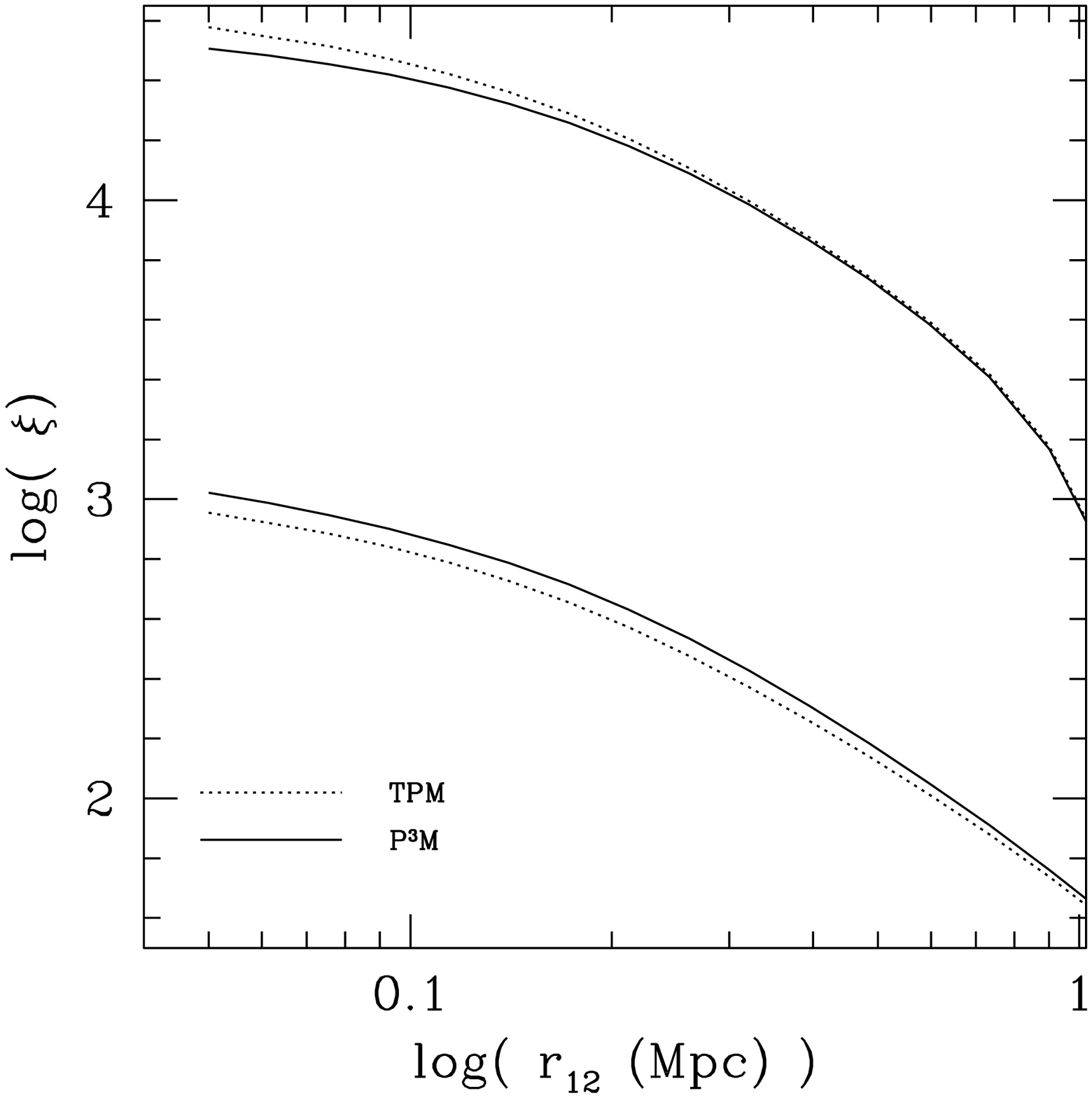}
\figcaption[f6.eps]{The particle--particle correlation function 
for halo particles from the TPM {\it (dotted line)}  and P$^3$M 
{\it (solid line)} simulations.  
The bottom pair of lines shows the correlation function for all
particles; the upper pair was calculated using only those 
particles in the 1000 most massive halos.
\label{figcorrel} }

\plotone{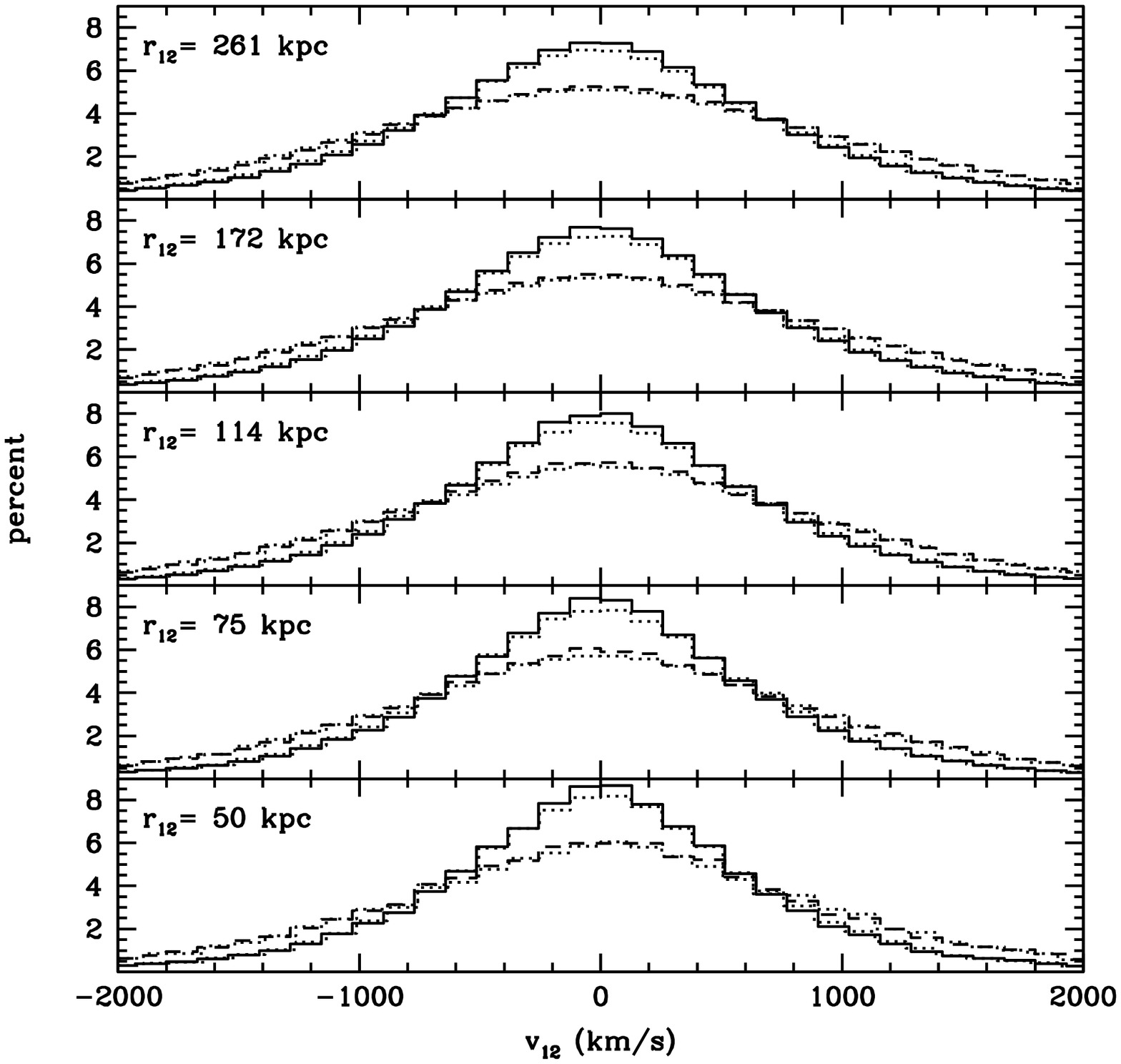}
\figcaption[f7.eps]{Histograms of the line-of-sight velocity
difference between pairs of particles with a given separation $r_{12}$.
{\it Solid lines:} Pairs from the 1000 most massive P$^3$M halos.
{\it Dashed lines:} Pairs from the 100 most massive P$^3$M halos.
{\it Dotted lines:} The corresponding values from the TPM simulation.
\label{figvhist} }

\plotone{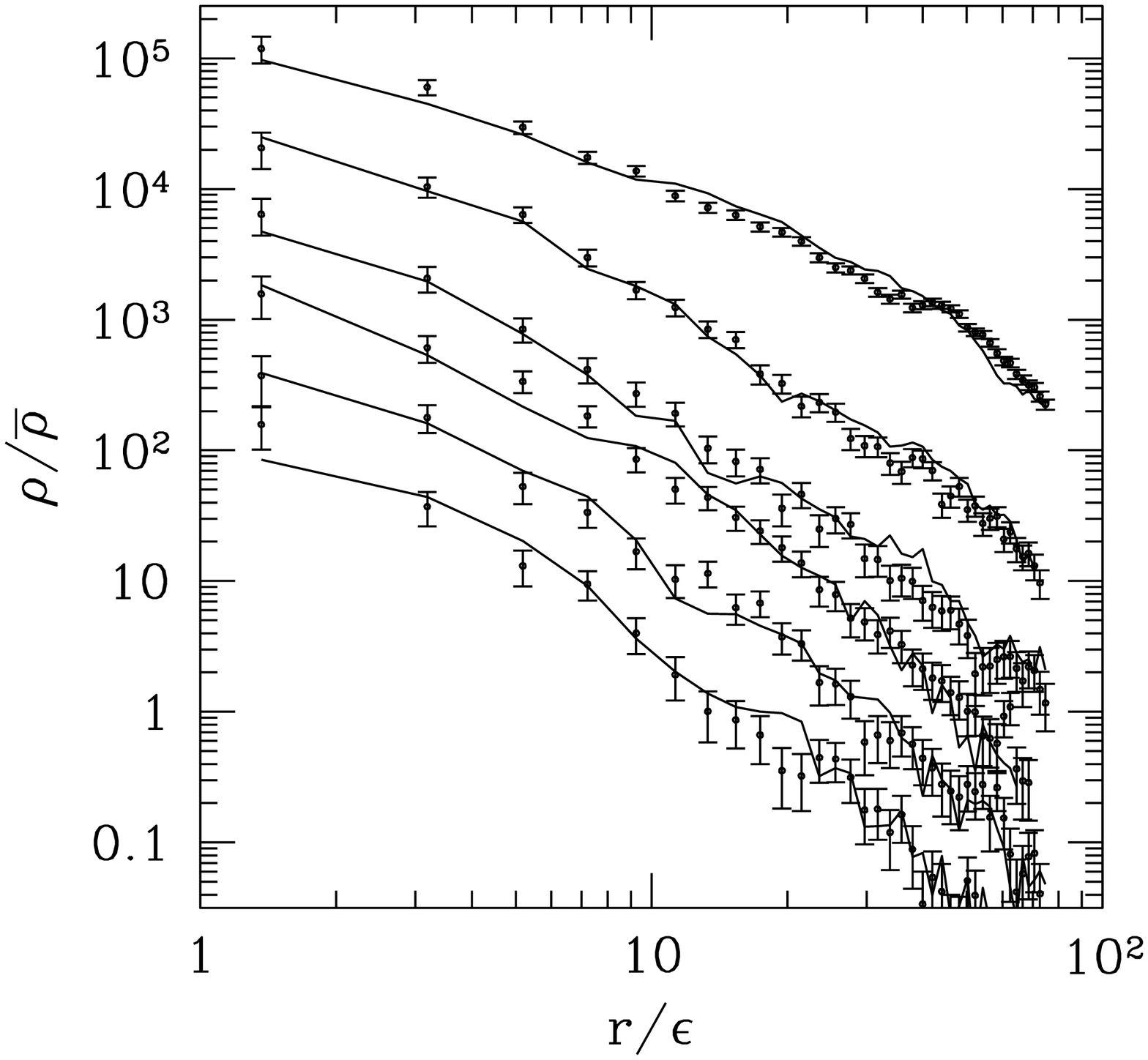}
\figcaption[f8.eps]{The density profile of halos, comparing
P$^3$M {\it (lines)} and TPM {\it (points)}. The error bars are the
square root of the number of particles in each radial bin. For clarity
each curve has been moved down by a factor of $\sqrt{10}$ from the
one above.  From top to bottom, the halos contain 6265, 1203, 531,
476, 319, and 173 particles.
\label{fighaloprof} }

\epsscale{0.85}
\plotone{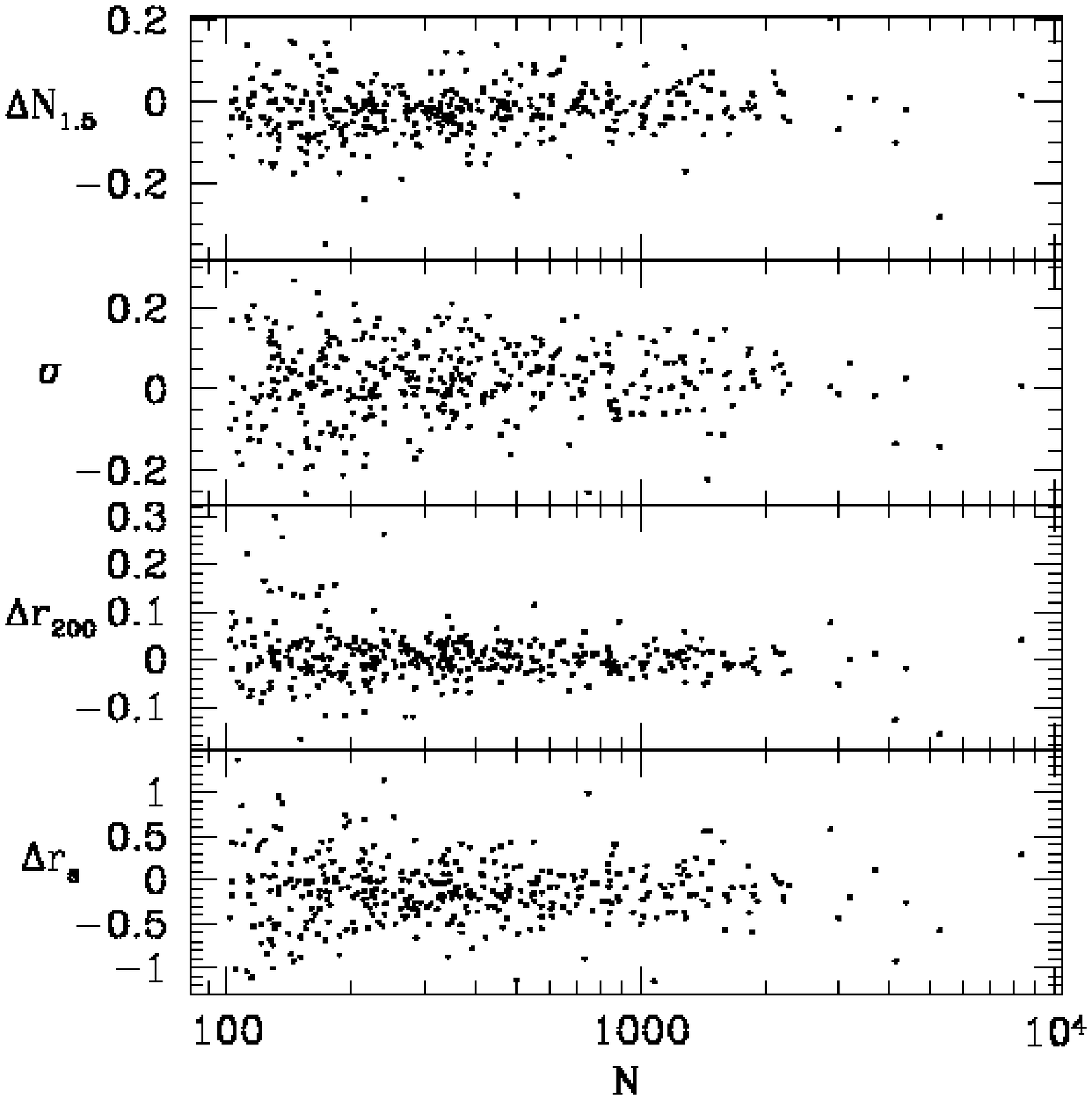}
\figcaption[f9.eps]{A comparison of TPM halo properties to P$^3$M.
Each panel show the fractional difference between the TPM halo and
the same halo in the P$^3$M simulation.
From top to bottom: the number of particles
within 1.5 Mpc/$h$ (1.76 TPM cells);  
the velocity dispersion;
and from a fit to the NFW profile, $r_{200}$ and $r_s$.
The $x$-axis is the number of particles in the halo according
to DENMAX.
\label{fighalocomp} }

\clearpage
\begin{deluxetable}{ccrrrrrrr}
\tablecolumns{9}
\tablecaption{Wallclock time in seconds on a 250 MHz SGI Origin 2000. \label{tbl-1}}
\tablehead{
\colhead{}  & \colhead{} & \multicolumn{3}{c}{z=9} &  \colhead{}  &
\multicolumn{3}{c}{z=0.5} \\ 
\cline{3-5} \cline{7-9} \\ 
\colhead{N} & \colhead{NPE} & 
\colhead{PM} & \colhead{DD} & \colhead{Tree} & \colhead{} & 
\colhead{PM} & \colhead{DD} & \colhead{Tree} }
\startdata

  $256^3$ & 4	&21.9 & 70.8 & 13.8	&  &19.3 & 117.1 & 2695.0 \\
  $256^3$ & 8	&11.1 & 35.8 & 7.0	&  &9.7 & 68.8 & 1350.0	 \\
  $256^3$ & 16	&5.6 & 18.6 & 3.8	&  &5.5 & 37.8 & 700.0	 \\
  $256^3$ & 32	&2.9 & 10.0 & 2.0	&  &3.4 & 160.1 & 339.5	 \\
  $256^3$ & 64	&2.0 & 6.8 & 1.0	&  &2.4 & 140.1 & 175.5	 \\
  $256^3$ & 128	&1.4 & 5.7 & 0.5	&  &2.0 & 234.0 & 84.5	 \\
  $256^3$ & 256	&1.6 & 11.7 & 0.2	&  &3.1 & 252.6 & 38.2	 \\
  \\
  $512^3$ & 32	&27.0 & 80.2 & 8.4	&   & 24.5 & 133.5 & 1085.0	 \\
  $512^3$ & 64	&13.8 & 48.5 & 4.2	&   & 14.4 & 67.2 & 545.0	 \\
  $512^3$ & 128	&7.6 & 33.4 & 2.1	&   & 9.9 & 44.7 & 275.5	 \\
  $512^3$ & 256	&13.6 & 38.4 & 1.1	&   & 12.9 & 38.8 & 144.5	 \\
  \\
  $1024^3$ & 256 &69.17 & 136.9 & 9.5	&   & 87.1 & 200.8 & 1433.0	 \\
\enddata
\tablecomments{ {\em PM:} The PM portion of the code.  {\em DD:} Time
spent preparing trees, including domain decomposition, tidal force
calculation, and any load imbalance.
{\em Tree:} The potential computation for tree particles. }
\end{deluxetable}


\begin{thebibliography}{DUM}
\bibitem[Bagla (1999)]{Bagla99} Bagla, J.S. 1999, preprint (astro-ph/9911025)
\bibitem[Barnes \& Hut 1986]{BarnHut86}Barnes, J. \& Hut, P. 1986, 
Nature, 324, 446
\bibitem[Bode \etal\ 1996]{BodeXuCe96}Bode, P., Xu, G. \& Cen, R. 1996,
    Supercomputing '96: Proceedings of the 1996 ACM/IEEE
    Supercomputing Conference, Pittsburgh: IEEE Computer Society
    (http://www.supercomp.org/sc96/)

\bibitem[Couchman 1991]{Couc91} Couchman, H.M.P. 1991, ApJ, 368, L23
\bibitem[Couchman (1997)]{Couc97} Couchman, H.M.P. 1997,
in Computational Astrophysics, ed. D.A. Clarke \& M.J. West,
(San Francisco: ASP), 340
\bibitem[Dav\'e et al. 1997]{DaveDubi97}Dav\'e, R.,  
    Dubinski, J. \& Hernquist, L. 1997, NewA 2, 277
\bibitem[Efstathiou et al. 1985]{Efst85}Efstathiou G., Davis, M.,
    Frenk, C. \& S. White 1985, ApJS, 57, 241
\bibitem[Ferrell \& Bertschinger 1994]{FerrBert94}Ferrell, R. \&
    Bertschinger, E.  1994, Int. J. Mod. Phys. C, 5, 933
\bibitem[Frederic 1997]{Fred97}Frederic, J.J. 1997, Ph.D. Thesis, MIT
\bibitem[Gelb \& Bertschinger 1994]{GelbBert94}Gelb, J.M. \&
    Bertschinger, E.  1994, ApJ, 436, 467
\bibitem[Hernquist 1987]{Hern87} Hernquist, L. 1987, ApJS, 64, 715
\bibitem[Hernquist \& Katz 1989]{HernKatz89} Hernquist, L. \& Katz, N.
    1989, ApJS, 70, 419
\bibitem[Hockney \& Eastwood 1981]{HockEast81}  Hockney, R.W. \& Eastwood, 
    J.W. 1981, Computer Simulation Using Particles (New York: McGraw Hill)
\bibitem[Knebe et al. (2000)]{Knebetal00} Knebe, A., Kravtsov, A.V.,
    Gottl\"ober, S. \& Klypin, A.A. 2000, preprint (astro-ph/9912257)
\bibitem[Kravtsov et al. 1997]{Kravetal97}  Kravtsov, A.V., Klypin, A.A.,
    \& Khokhlov, A.M. 1997, ApJS, 111, 73
\bibitem[NFW]{NFW97}Navarro, J.F., Frenk, C.S. \& White, S.D.M. 1997,
    ApJ, 490, 493
\bibitem[Melott et al. (1997)]{Meloetal1997} Melott, A.L., Splinter, R.J.,
    Shandarin, S.F. \& Suto, Y. 1997, ApJ 479, L79
\bibitem[Norman \& Bryan 1999]{NormBrya99} Norman, M.L. \& Bryan, G.L.
    1999, in Numerical Astrophysics, ed. S. Miyama, K. Tomisaka \& T. Hanawa, 
    (Dordrecht: Kluwer Academic), 19
\bibitem[Ostriker \& Steinhardt (1995)]{OstrStei95} Ostriker, J.P. \&
Steinhardt, P.J. 1995, Nature, 377, 600
\bibitem[Pearce \& Couchman 1997]{PearCouc97}Pearce, F.R. \& Couchman,
    H.M.P. 1997, NewA 2, 411
\bibitem[Splinter et al. (1998)]{Splinetal1998} Splinter, R.J., Melott, A.L.,
   Shandarin, S.F. \& Suto, Y. 1998, ApJ, 497, 38
\bibitem[Xu (1995)]{Xu95}Xu, G. 1995, ApJS, 98, 355
\end{thebibliography}
\end{document}